\documentclass{article}

\pdfoutput=1

\usepackage[utf8]{inputenc}
\usepackage{authblk}
\usepackage[colorlinks]{hyperref}
\usepackage{breakurl}
\usepackage{amsmath}
\usepackage{amssymb}
\usepackage{fancyhdr}% http://ctan.org/pkg/fancyhdr
\usepackage{graphicx}

\title{
How does noise protection affect the accuracy of life expectancy and other demographic indicators?
}
\author[1]{Fabian Bach
    \thanks{\href{mailto:Fabian.BACH@ec.europa.eu}{Fabian.BACH@ec.europa.eu}}
    }
\affil[1]{European Commission (Eurostat), 2920 Luxembourg, Luxembourg
    \thanks{% EC disclaimer
    The views expressed are purely those of the author and may not in any circumstances be regarded as stating an official position of the European Commission.}
    }

\date{\today}

\usepackage{natbib}
\usepackage{graphicx}

\begin{document}

\maketitle

\begin{abstract}
    New and efficient methods based on noise addition to protect the confidentiality in population statistics have been developed, tested and applied in census production by various members of the European Statistical System over the past years.  Basic demographic statistics---such as population stocks, live births and deaths by age, sex and region---may be protected in a similar way, but also form the raw input to calculate various demographic indicators.  This paper analyses the impact on the accuracy of some selected indicators, namely fertility and mortality rates and life expectancies, under the assumption that the raw input counts are protected with a generic noise method with fixed variance parameter, by comparing the size of noise uncertainties with intrinsic statistical uncertainties using a Poisson model.  As a by-product, we derive and validate numerically a closed analytical expression for the variance of life expectancies in a certain class of calculation models as a function of the variance of input mortality data.  This expression also allows to calculate analytically the statistical uncertainty of life expectancies using the mentioned Poisson model for the input death counts.
\end{abstract}

\section{Introduction}\label{}

After the 2011 EU census, European Statistical System (ESS)\footnote{The joint body of Eurostat and the national statistical institutes of all EU countries and Iceland, Liechtenstein, Norway and Switzerland. It is responsible for the development and quality assurance of social European statistics.}
experts realised that traditional statistical disclosure control (SDC) methods like cell suppression become inefficient when applied to very detailed cross-tabulations.  Moreover, these methods do not protect outputs effectively when disseminated on non-nested geographical breakdowns like administrative regions and grids (geographic differencing)~\cite{essnet2017}.  Mainly for these two reasons, ESS efforts began from 2015 onwards to explore new SDC methods that avoid excessive information loss and also protect against geographic differencing.  As a result, two methods based on adding some noise to the statistics were tested and recommended for the 2021 EU census~\citep{essnet2017,essnet2019,essnet2019a}, where a significant number of ESS members eventually decided to use one or a combination of these methods for their census outputs.

In the wake of the modernisation of social statistics in the ESS, it is likely that also the annual demographic statistics would become more detailed and eventually integrated with multi-annual population statistics like censuses.  This would entail some new questions, e.g.\ whether some of these annual demographic statistics would require confidentiality protection; how the consistency with census-like (protected) statistics would be ensured; and finally how the key use cases of demographic statistics can be sufficiently ensured (cf.~\citep{stace2024}).

For instance, basic demographic statistics including population stocks, live births and deaths by age, sex and region are needed to calculate various demographic indicators at regional level.  If the protection methods originally developed for census outputs were to be applied to these demographic statistics, the question arises how the additional noise-induced uncertainty propagates to the resulting indicator values.  This paper analyses the effects of a generic noise protection method with a fixed variance parameter on some selected demographic indicators.

\section{Noise setup and uncertainty propagation}\label{setup}

It is one of the features of the {\it cell key method} (CKM)~\citep{fraser2005,marley2011,thompson2013}---one of the two methods recommended for census ouputs in the ESS---that the variance of the noise added to the output counts is a fixed input parameter, here denoted~$V$.  Moreover, the detailed noise distributions of the CKM typically follow a discretised bell curve (as all noise terms must be integers to retain integer outputs) that is very close to a Gaussian~\citep{giessing2016,bach2021}.
Through these properties, the CKM is considered a reference example for a generic noise protection method in the sense mentioned above.

Furthermore, the standard theory of error propagation based on linear response of a measurement function~$f(\mathbf{x})$ to a set of uncorrelated input measurements~$\mathbf{x}$ can be applied:
\begin{equation}\label{eq_errprop}
    \Delta_f = \sqrt{ \sum_i \left( \frac{\partial f}{\partial x_i} \right)^2 \Delta_i^2 } \;,
\end{equation}
where $i$ runs over the elements of $\mathbf{x}$ and $\Delta$ denotes the standard deviation. 
From Eq.~\ref{eq_errprop} it is straightforward to derive two simple but handy special cases: when $f$ is a simple sum of independent input measurements, the squares of the absolute uncertainties are added; and when $f$ is a simple product (or ratio) of independent input measurements, the squares of relative uncertainties are added---i.e.
\begin{subequations}
\begin{align}%\label{eq_errprop_simple}
%    \begin{align}
        f\left(\mathbf{x}\right) = \sum_i x_i \quad  &\Rightarrow\quad \Delta_f = \sqrt{\sum_i \Delta_i^2} \label{eq_errprop_sum}\\
        f\left(\mathbf{x}\right) = \prod_i x_i^{\pm 1} \quad &\Rightarrow\quad \delta_f = \sqrt{\sum_i \delta_i^2}\label{eq_errprop_prod}
%    \end{align}
\end{align}
\end{subequations}
with relative uncertainties of the input measurements denoted $\delta_i = \Delta_i/x_i$.

As a further simplification, in the specific setup analysed here where only the CKM effect is to be considered, the standard deviation of all the input measurements~$x_i$ is constant:
\begin{equation}\label{eq_delta_simple}
    \Delta_i \equiv \Delta = \sqrt{V} \quad \forall i.
\end{equation}

\section{Results for selected demographic indicators}\label{calc}

For the quantitative analysis we select three demographic indicators that are generally very important for all kinds of demographic analysis: crude fertility and death rates, and life expectancy.  These indicators depend on three distinct sets of raw input counts for a given region and time period: population stocks by sex and single years of age at the beginning and at the end of a given time period (typically a calendar year), as well as the number of live births (by mother's single year of age) and the number of deaths (by sex and single year of age) during the same time period.  To add some geographical granularity, we analyse all NUTS~2 regions of all ESS members for the latest available reference year 2023, as published in the data sets {\it demo\_r\_d2jan}\footnote{\href{https://ec.europa.eu/eurostat/databrowser/view/DEMO_R_D2JAN/default}{\url{https://ec.europa.eu/eurostat/databrowser/view/DEMO_R_D2JAN/default}}}, {\it demo\_r\_fagec}\footnote{\href{https://ec.europa.eu/eurostat/databrowser/view/DEMO_R_FAGEC/default}{\url{https://ec.europa.eu/eurostat/databrowser/view/DEMO_R_FAGEC/default}}} and {\it demo\_r\_magec}\footnote{\href{https://ec.europa.eu/eurostat/databrowser/view/DEMO_R_MAGEC/default}{\url{https://ec.europa.eu/eurostat/databrowser/view/DEMO_R_MAGEC/default}}}.

\subsection{Fertility rates}\label{calc_fert}

The age-specific crude fertility rate is defined as
\begin{equation}\label{eq_def_fx}
    f_x = \frac{b_x}{w_x}
\end{equation}
with $x$ the relevant age in single years, $b_x$ the number of live births by mothers aged~$x$ and $w_x$ the average female population stock aged~$x$ during the reference time period.
Then the relative error on $f_x$ is given by Eq.~\ref{eq_errprop_prod} as
\begin{equation}\label{eq_delta_fx}
    \delta_{f_x} \equiv \frac{\Delta_{f_x}}{f_x} = \Delta \sqrt{\frac{1}{b_x^2} + \frac{1}{w_x^2}} \simeq \frac{\Delta}{b_x}\;,
\end{equation}
where in the last step we neglect for simplicity a possible CKM uncertainty on~$w_x$, because $b_x \ll w_x$ in general.
In this approximation, the absolute uncertainty on $f_x$ from Eq.~\ref{eq_delta_fx} is $\Delta_{f_x}=f_x\delta_{f_x}\simeq\Delta/w_x$.

\begin{figure}[t]
\centering
\includegraphics[trim={0 60 0 35},scale=0.5]{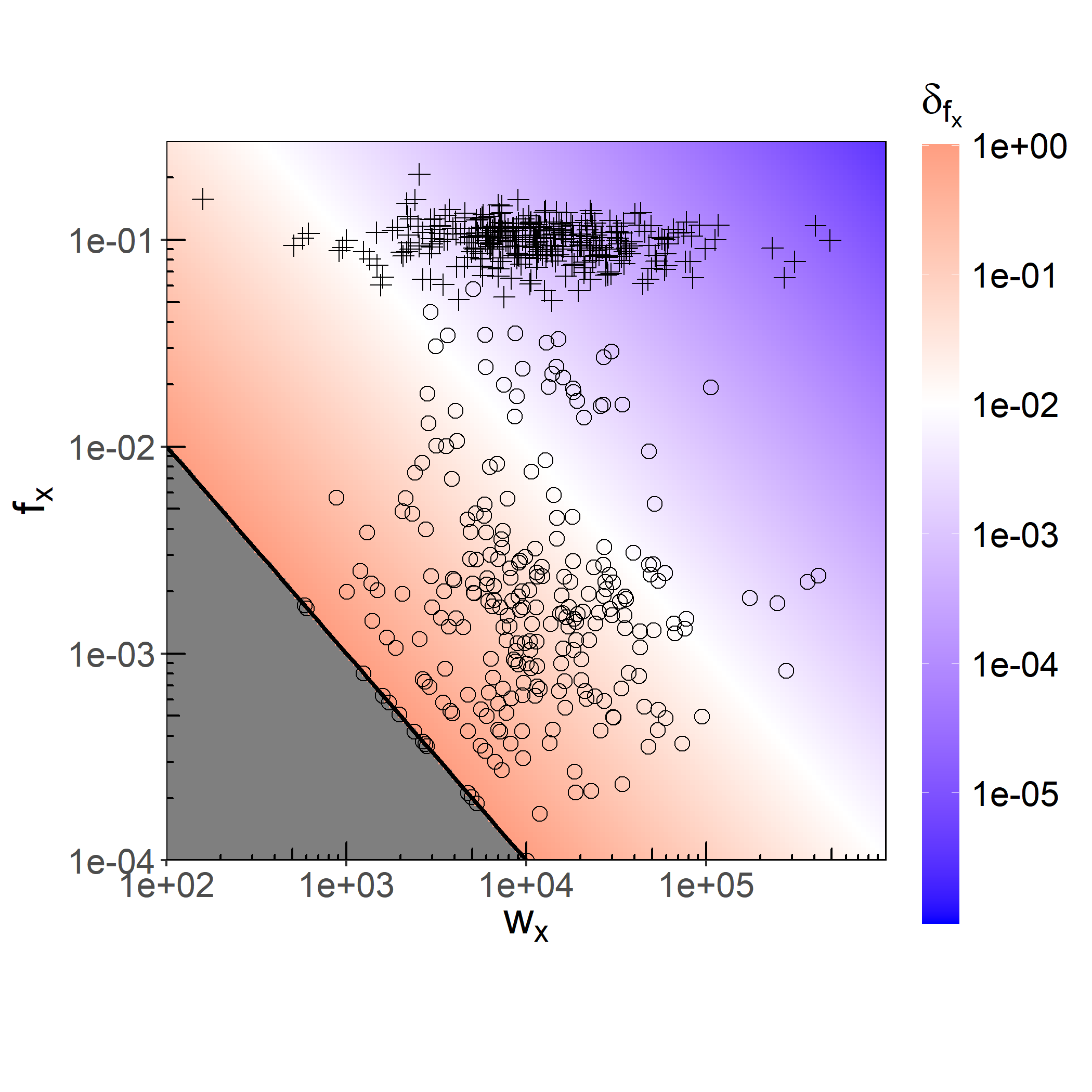}
\caption{Relative uncertainty on the age-specific fertility rate $\delta_{f_x}$ from CKM noise with $V=1$ as a function of $w_x$ and $f_x$ according to Eq.~\ref{eq_delta_fx}.  The black line indicates the limiting contour corresponding to $b_x=1$.  Actual fertility rates found in the 2023 NUTS~2 data are shown for $x=16$ (circles) and $x=30$ (cross-hairs), showing that most $\delta_{f_{16}}>1\%$ up to $100\%$ while most $\delta_{f_{30}}<1\%$.}
\label{fig_fx_CKM}
\end{figure}

Setting $V=1$ and plugging in the numbers from the data sets, one finds that in the majority of cases the relative CKM uncertainty on the age-specific fertility rates is moderate, with a median relative uncertainty of $0.37\%$ and more than $88\%$ of all relative uncertainties ($10\,854$ of $12\,287$ calculated) $<10\%$. Nevertheless, $33\%$ of the uncertainties found ($4\,061$ of $12\,287$) are $>1\%$.  In line with Eq.~\ref{eq_delta_fx}, it is not surprising that the largest relative uncertainties, up to $100\%$ for $V=1$, are found where the numbers of live births are very small, i.e.\ $b_x<10$ down to $b_x=1$. This is illustrated in Fig.~\ref{fig_fx_CKM} showing the parametric plot of Eq.~\eqref{eq_delta_fx} over the entire relevant $w_x$--$f_x$ plane.

Finally, one can also analyse the total fertility rate given as
\begin{equation}\label{eq_def_f}
    f = \sum_x f_x\;,
\end{equation}
so that by applying Eq.~\ref{eq_errprop_sum} and reusing $\Delta_{f_x}$, the relative uncertainty on $f$ amounts to
\begin{equation}\label{eq_delta_f}
    \delta_f = \frac{\Delta_f}{f} = \frac{\Delta}{f} \sqrt{\sum_x \frac{1}{w_x^2}}\;.
\end{equation}
After plugging in the numbers, there is again little surprise that in this more aggregate indicator the CKM uncertainty is negligible in basically all cases calculated: more than $83\%$ of all $\delta_f$ calculated ($294$ of $356$) are $<0.1\%$, and $\delta_f>1\%$ only for the smallest NUTS~2 region in the entire EU: \r{A}land (FI20).

\subsection{Mortality rates}\label{calc_mort}
\begin{figure}[t]
\centering
\includegraphics[trim={0 60 0 35},clip,scale=0.5]{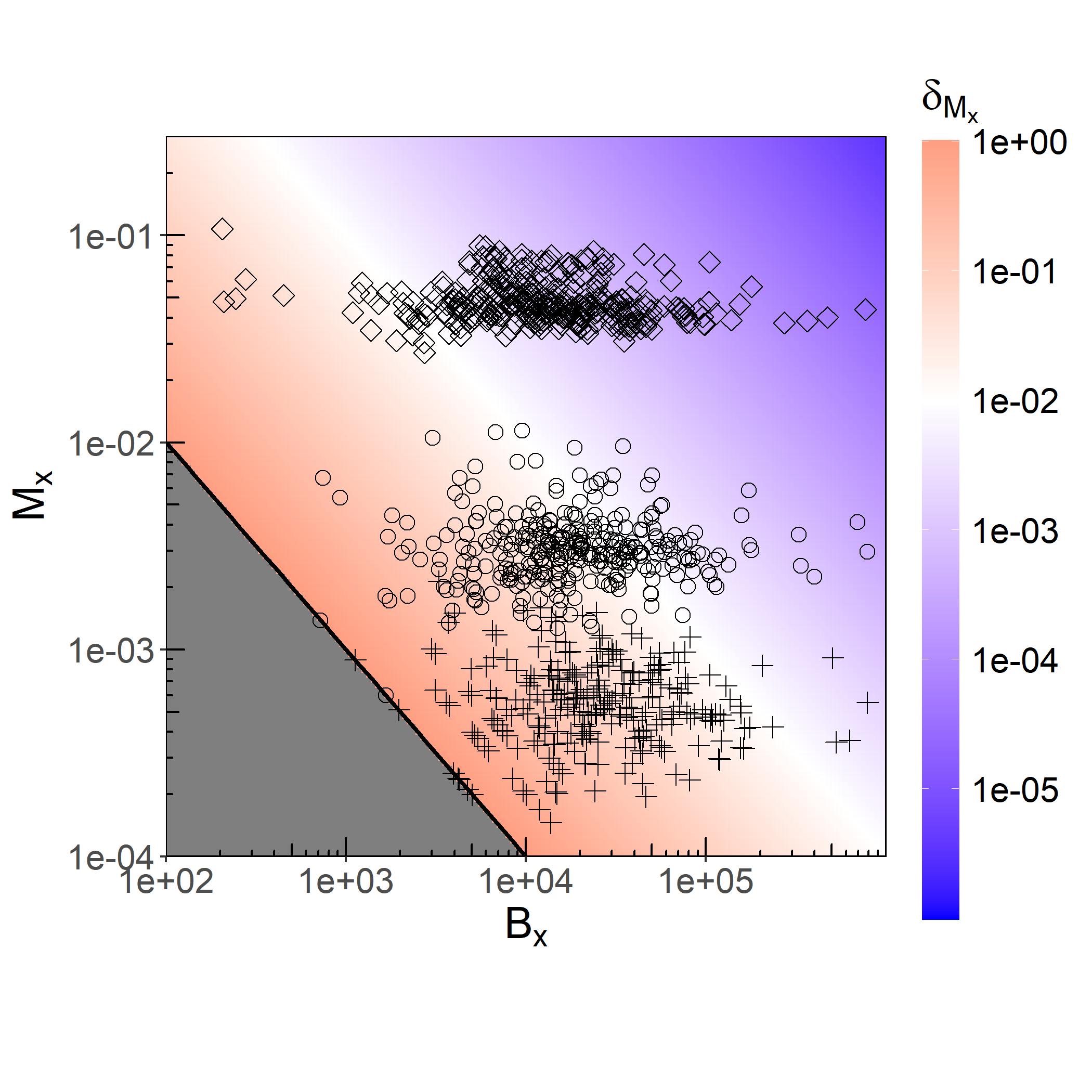}
\caption{Relative uncertainty on the age-specific mortality rate $\delta_{M_x}$ from CKM noise with $V=1$ as a function of $B_x$ and $M_x$ according to Eq.~\ref{eq_delta_Mx}.  The black line indicates the limiting contour corresponding to $D_x=1$.  Actual mortality rates found in the 2023 NUTS~2 data are shown for $x=0$ (circles), $x=30$ (cross-hairs) and $x=80$ (diamonds), showing that most $\delta_{M_{0,30}}>1\%$ up to $100\%$ while most $\delta_{M_{80}}<1\%$.}
\label{fig_Mx_CKM}
\end{figure}

The age-specific crude mortality rate is defined as
\begin{equation}\label{eq_def_Mx}
    M_x = \frac{D_x}{B_x}
\end{equation}
with $D_x$ the number of deaths at age~$x$ and $B_x$ the average population during the reference time period.  These rates can be calculated for the total population and for the male and female populations, respectively. Analogously to Eq.~\ref{eq_delta_fx} with $D_x\ll B_x$, the relative uncertainty is then
\begin{equation}\label{eq_delta_Mx}
    \delta_{M_x} \simeq \delta_{D_x} = \frac{\Delta}{D_x}\; .
\end{equation}
Plugging in the NUTS~2 numbers from 2023 with $V=1$, one finds qualitatively similar results to section~\ref{calc_fert}. However, quantitatively the median relative uncertainty is $1.4\%$ for the total population across all ages ($2.2\%$ for male and $2.9\%$ for female populations), and in the total population $23\%$ of all uncertainties calculated are $>10\%$ ($27\%$ respectively $33\%$ for males and females).  This is because the small-number effect already discussed in section~\ref{calc_fert} is much more pronounced in the age-specific mortality rates:  deaths are spread across all ages instead of just a fertile age band (e.g.\ 16 to 44 years), and they are (fortunately) very scarce across a majority of ages, increasing only in the terminal ages.  Therefore, age-specific death rates stemming from very small death counts $D_x<10$ are much more numerous than age-specific fertility rates, especially at the NUTS~2 level considered here, leading to the larger typical relative uncertainties found above.  As for the fertility rates, Fig.~\ref{fig_Mx_CKM} shows the parametric dependence of $\delta_{M_x}(B_x,M_x)$ according to Eq.~\eqref{eq_delta_Mx} over the entire relevant parameter space and showing some actual data points from the 2023 NUTS~2 data.

\subsection{Life expectancy}\label{calc_lexp}

The calculation of age-specific life expectancy $E_x$ is considerably more complex than that of the crude rates presented in the previous sections.  Therefore, also the uncertainty propagation according to Eq.~\ref{eq_errprop} is more complex and it is worthwhile to use the auxiliary indicators published in the so-called {\it life table}, see data set {\it demo\_r\_mlife}\footnote{\href{https://ec.europa.eu/eurostat/databrowser/view/DEMO_R_MLIFE/default}{\url{https://ec.europa.eu/eurostat/databrowser/view/DEMO_R_MLIFE/default}}; note that German NUTS~2 regions are not analysed as they are not available in the public dataset.\label{fn_demo_r_mlife}} for the ESS.  Most notably, we make use of the total number of person-years lived from age $x$ until all have died ($T_x$), the total number of person-years lived from age $x$ to $x+1$ ($L_x$), the survivorship function ($\ell_x$) and the age-specific mortality rate ($M_x$), all included in the life table.  The sole raw data input is through $M_x$ depending on the raw number of age-specific deaths~$D_x$ and the average population of that age~$B_x$ for the given period.  All other life table indicators and ultimately $E_x$ are derived from this as
%follows\footnote{See e.g.\ the illustrative explanation of relations between life table indicators on \href{https://www.lifeexpectancy.org/lifetable.shtml}{\url{https://www.lifeexpectancy.org/lifetable.shtml}}.\label{fn_ltable}}:
follows\footnote{Cf.\ e.g.\ Eurostat's life expectancy methodology documented at \url{https://ec.europa.eu/eurostat/cache/metadata/Annexes/demo_mor_esms_an_1.pdf}.\label{fn_ltable}}:
\begin{equation}\label{eq_def_Ex}
\begin{aligned}
    E_x = \frac{T_x}{\ell_x} = \frac{1}{\ell_x}\sum_{k=x}^{\overline{x}} L_k =& %\frac{\ell_x/2+\sum_{k=x+1}^{\overline{x}}\ell_k}{\ell_x} \\
    \frac{1}{2} + \sum_{k=x+1}^{\overline{x}-1}\frac{\ell_k}{\ell_x} + \left(\frac{1}{2}+\frac{1}{M_{\overline{x}}}\right)\frac{\ell_{\overline{x}}}{\ell_x} \\
    &\quad\text{with}\quad L_{x<{\overline{x}}} = \frac{\ell_x + \ell_{x+1}}{2}\quad\text{and}\quad L_{\overline{x}}=\frac{\ell_{\overline{x}}}{M_{\overline{x}}}\;,
\end{aligned}
\end{equation}
where the sums run from the age~$x$ in question until the open-ended terminal age class~$\overline{x}$ where the age-specific tabulation stops.\footnote{The cut-off age depends on national publication practices and other constraints; e.g.\ in the EU life tables (footnote~\ref{fn_demo_r_mlife}) this terminal age class is `85 years or over'.\label{fn_xbar}}
Note the explicit dependency on the terminal mortality rate $M_{\overline{x}}$ entering through the terminal person-years lived $L_{\overline{x}}$.  In general, there is some freedom in modelling $L_{\overline{x}}$ as a function of $M_{\overline{x}}$ (cf.\ e.g.\ footnote~\ref{fn_ltable}), while the relation above is the one used by Eurostat. 
A further implicit dependency on all other mortality rates~$M_{x<\overline{x}}$ enters through the $\ell_x$ which are here defined recursively as\footnote{The attentive reader may have noted the difference between the definitions of the survivorship function in the Eurostat methodology (footnote~\ref{fn_ltable}) and here in Eq.~\eqref{eq_def_lx}: The two definitions have the same Taylor expansion in $M_x$ down to $\mathcal{O}(M_x^3)$. Since prevalently $M_x\ll 1$ (cf.\ Section~\ref{calc_mort}), the definitions are therefore equivalent for all practical purposes, while the exponential definition of Eq.~\eqref{eq_def_lx} allows for the following analytic manipulations necessary to derive Eq.~\eqref{eq_delta_Ex_final}.  The exponential definition is used e.g.\ in the U.S., see \url{https://www.lifeexpectancy.org/lifetable.shtml}.}
\begin{equation}\label{eq_def_lx}
    \ell_{x+1} = \ell_x \exp\left(-M_x\right) \quad\text{with}\quad \ell_o\equiv 100\,000\;,
\end{equation}
so that
\begin{equation}\label{eq_lxn}
    \frac{\ell_{x+n}}{\ell_x} = \exp\left( -\sum_{k=x}^{x+n-1} M_k \right).
\end{equation}
This can be plugged into Eq.~\ref{eq_def_Ex} to replace all $\ell$~ratios, giving
\begin{equation}\label{eq_Ex_Dx}
    E_x(M_k) = \frac{1}{2} + \sum_{n=x+1}^{\overline{x}-1} \exp\left( -\sum_{k=x}^{n-1} M_k \right) + \left(\frac{1}{2}+\frac{1}{M_{\overline{x}}}\right) \exp\left( -\sum_{k=x}^{\overline{x}-1} M_k \right).
\end{equation}
Finally, to derive the uncertainty on $E_x$, one needs the functional dependency of $M_k$ on $D_k$ and $B_k$.  While again there is some methodological freedom in modelling the mortality rates for life expectancy 
purposes\footnote{
%E.g.\ the UK's Office for National Statistics averages over the three most recent years (\href{https://www.ons.gov.uk/peoplepopulationandcommunity/healthandsocialcare/healthandlifeexpectancies/methodologies/guidetocalculatingnationallifetables}{\url{https://www.ons.gov.uk/peoplepopulationandcommunity/healthandsocialcare/healthandlifeexpectancies/methodologies/guidetocalculatingnationallifetables}}), or Eurostat applies Farr's death rate method for the European core health indicators (\href{https://ec.europa.eu/health/indicators/docs/echi_10_ds_en.pdf}{\url{https://ec.europa.eu/health/indicators/docs/echi_10_ds_en.pdf}}).\label{fn_Mx}}, 
E.g.\ the UK's Office for National Statistics averages over the three most recent years, see \href{https://www.ons.gov.uk/peoplepopulationandcommunity/healthandsocialcare/healthandlifeexpectancies/methodologies/guidetocalculatingnationallifetables}{\url{https://www.ons.gov.uk/peoplepopulationandcommunity/healthandsocialcare/healthandlifeexpectancies/methodologies/guidetocalculatingnationallifetables}}.\label{fn_Mx}}, 
we assume here for practical purposes the simplest model---also used by Eurostat (see footnote~\ref{fn_ltable})---that employs the crude age-specific mortality rate of Eq.~\ref{eq_def_Mx}, i.e.\ $M_k=D_k/B_k$.
Ignoring again any potential noise on $B_k$, as argued in Section~\ref{calc_fert}, Eq.~\eqref{eq_Ex_Dx} can now be differentiated with respect to the raw input~$D_k$ using the chain rule to calculate the resulting uncertainty on $E_x$ as
\begin{equation}\label{eq_delta_Ex}
    \delta_{E_x} = \frac{\Delta}{E_x}\sqrt{\sum_{z=x}^{\overline{x}} \left( \frac{\partial E_x}{\partial M_z} \right)^2\left( \frac{\partial M_z}{\partial D_z} \right)^2} \quad\text{with}\quad \frac{\partial M_z}{\partial D_z} = \frac{1}{B_z}
\end{equation}
and\footnote{To see the following, rewrite the exponential of a sum as a product of exponentials, i.e.\ $\exp(\sum_k\alpha_k)=\prod_k A_k$ with $A_k=\exp(\alpha_k)$, and then expand the sum over products of exponentials as $A_1 + A_1 A_2 + \cdots + A_1 A_2\cdots A_{z} + A_1 A_2\cdots A_z A_{z+1} +\cdots$. The derivative of this with respect to index~$z$ has the form $(A_1 A_2\cdots A_z)(1+A_{z+1}+\cdots)$.}
\begin{subequations}
\begin{align}
    \frac{\partial E_x}{\partial M_{z<\overline{x}}} =& \frac{\partial}{\partial M_{z<\overline{x}}} \left[\sum_{n=x+1}^{\overline{x}-1} \exp\left( -\sum_{k=x}^{n-1} M_k \right)+ \left(\frac{1}{2}+\frac{1}{M_{\overline{x}}}\right) \exp\left( -\sum_{k=x}^{\overline{x}-1} M_k \right)\right] \nonumber\\
    =& -\underbrace{\exp\left( -\sum_{k=x}^{z} M_k \right)}_{=\frac{\ell_{z+1}}{\ell_x}}\times \nonumber\\
    &\times\underbrace{\left[1 + \sum_{n=z+1}^{\overline{x}-2}\exp\left( -\sum_{k=z+1}^{n} M_k \right) + \left(\frac{1}{2}+\frac{1}{M_{\overline{x}}}\right) \exp\left( -\sum_{k=z+1}^{\overline{x}-1} M_k \right)\right]}_{=\frac{1}{2}+E_{z+1}}\;;\label{eq_dEx_dMz}\\
    \frac{\partial E_x}{\partial M_{\overline{x}}} =& -\frac{1}{M_{\overline{x}}^2} \frac{\ell_{\overline{x}}}{\ell_x}\;, \label{eq_dEx_dMxbar}
\end{align}
\end{subequations}
where Eqs.~\eqref{eq_lxn} and~\eqref{eq_Ex_Dx} were used in Eq.~\eqref{eq_dEx_dMz} to identify expressions in terms of $\ell_{z+1}$ and $E_{z+1}$.
This can be plugged back into Eq.~\ref{eq_delta_Ex} to obtain the final formula for the relative uncertainty\footnote{Since the analytical derivation of this formula is not straightforward, its correctness is checked numerically in Annex~\ref{Delta_Ex_check}.\label{fn_Delta_Ex_check}}
\begin{equation}\label{eq_delta_Ex_final}
    \delta_{E_x} = \frac{\Delta}{E_x}\sqrt{\sum_{z=x}^{\overline{x}-1} \frac{1}{B_z^2}\left(\frac{\ell_{z+1}}{\ell_x}\right)^2\left( \frac{1}{2}+E_{z+1} \right)^2 + \frac{B_{\overline{x}}^2}{D_{\overline{x}}^4} \left(\frac{\ell_{\overline{x}}}{\ell_x}\right)^2}\;.
\end{equation}
Since the analytical derivation of this result via Eqs.~\eqref{eq_Ex_Dx}, \eqref{eq_delta_Ex}, \eqref{eq_dEx_dMz} and~\eqref{eq_dEx_dMxbar} is not straightforward, its correctness is validated numerically in Annex~\ref{Delta_Ex_check}.
Eq.~\eqref{eq_delta_Ex_final} represents a closed analytical expression for $\delta_{E_x}$ for a class of life expectancy calculation methods characterised by Eqs.~\eqref{eq_def_lx} and~\eqref{eq_def_Mx}, which includes the EU method. It allows the straightforward calculation of $\delta_{E_x}$ from the CKM setup (recall $\Delta=\sqrt{V}$), from the life table indicators $\ell_x$ and $E_x$, and from the population stocks $B_x$ and the terminal death number~$D_{\overline{x}}$.

\begin{figure}[t]
\centering
\includegraphics[trim={0 60 0 60},clip,scale=0.5]{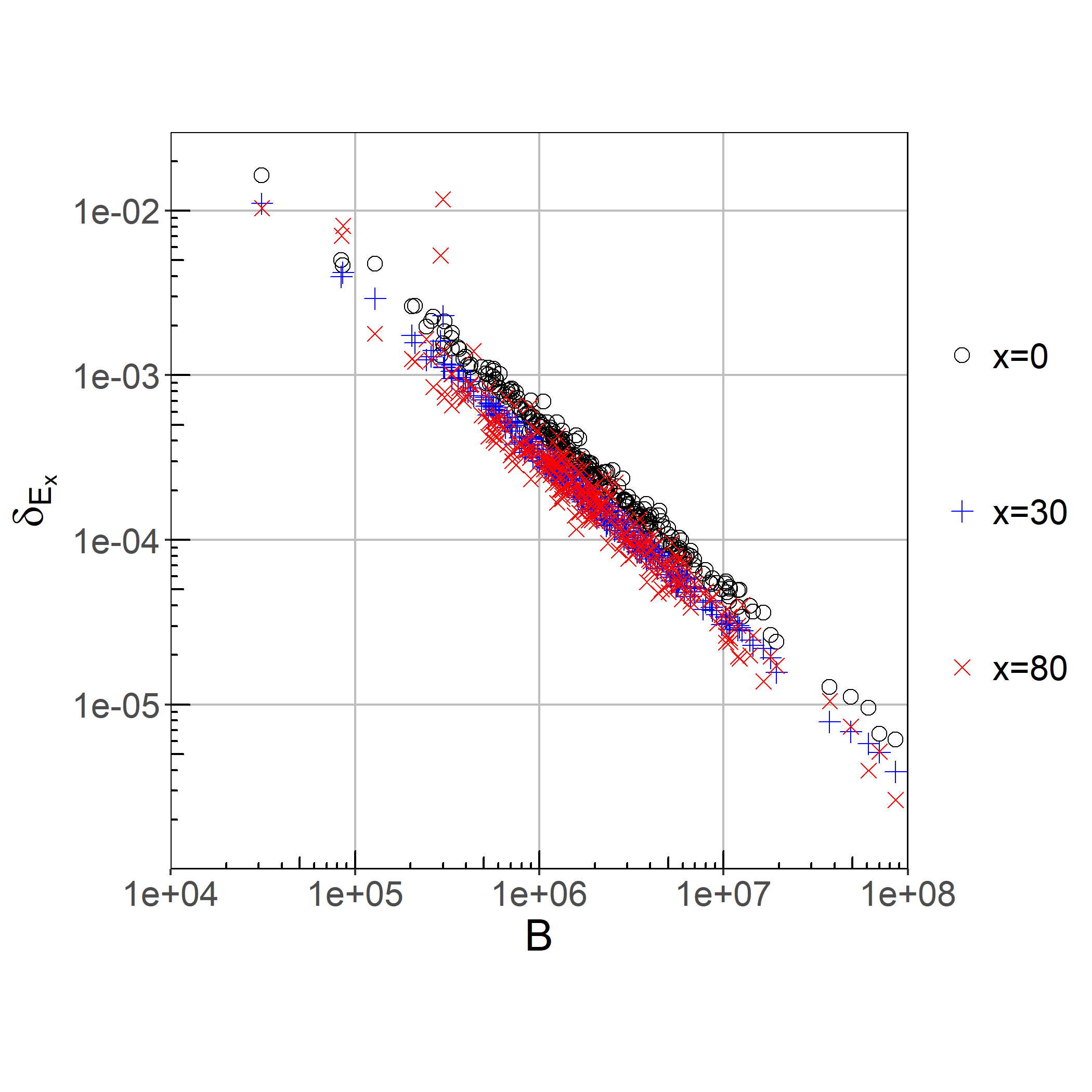}
\caption{Scatter plot of the relative uncertainty from CKM noise on the life expectancy at ages $x=0$, 30 and 80, according to Eq.~\eqref{eq_delta_Ex_final}, over the total population~$B$, with one dot for each ESS region analysed.}
\label{fig_Ex_CKM}
\end{figure}

Plugging in the 2023 numbers for NUTS~2 regions published by Eurostat with $V=1$, one finds very moderate relative uncertainties on the age-specific life expectancies largely prevailing, with a median $\delta_{E_x}=0.018\%$ for the total population ($0.037\%$ for the male population and $0.035\%$ for the female population), and $\delta_{E_x}>1\%$ for the total population occurring only in two NUTS~2 regions: Mayotte (FRY5) and again the smallest one, \r{A}land (FI20); see the corresponding data points in Fig.~\ref{fig_Ex_CKM} above the $\delta_{E_x}=1\%$ line at $B\simeq 3\times 10^5$ (Mayotte) and at $B\simeq 3\times 10^4$ (\r{A}land).
%However, while these numbers appear quite low and even negligible for the NUTS~2 level analysis here, 
Furthermore, Eq.~\ref{eq_delta_Ex_final}
%scales roughly with the total (male or female) population of a given region as
suggests an inverse correlation $\delta_{E_x}\sim \overline{B}_x$
with the total (male or female) population aged $\geq x$ of a given region
%as
%\begin{equation}\label{eq_delta_Ex_B_scaling}
%    \Delta_{E_x}\sim \frac{1}{B} \quad\text{with}\quad B=\sum_x B_x\;.
%\end{equation}
$\overline{B}_x=\sum_{k=x}^{\overline{x}} B_k$.
This is also illustrated in Fig.~\ref{fig_Ex_CKM} for various age bands~$x$.
%\begin{figure}[h!]
%\centering
%\includegraphics[scale=0.4]{fig/Scatterplot_B_E0.png}
%\caption{A scatter plot of the relative uncertainty on the life expectancy at birth $\delta_{E_0}$ over the total population~$B$, with one dot for each ESS region analysed.}
%\label{fig_dEx_B}
%\end{figure}

%A model based on Eq.~\ref{eq_delta_Ex_B_scaling} fitted to all available data points gives the coefficients
%\begin{equation}\label{eq_delta_Ex_B_fit}
%    \delta_{E_x}(B) \simeq 123 + \frac{123}{B}\;.
%\end{equation}
%By extrapolating this fitted model, 
%By fitting and extrapolating the correlation as indicated in Fig.~\ref{fig_dEx_B},
%one must assume that the relative noise impact on the uncertainty may increase to become significant in regions with small populations. At NUTS~2 level, \r{A}land with ca.\ $30\,000$ people is an absolute exception, but at NUTS~3 level there are roughly 20~regions of similar or smaller size with $\lesssim 40\,000$ people. In particular, the relative uncertainty would enter a regime of $\delta_{E_x}\gtrsim 5\%$ for the few NUTS~3 regions with $\lesssim 15\,000$ people and $V=1$ (with relative errors scaling as $\delta\sim\sqrt{V}$ according to Eqs.\ \eqref{eq_errprop} and~\eqref{eq_delta_simple}).  Naturally, the situation is worse with sex-specific life expectancies, where the underlying population sizes are even smaller.  Note that the same reasoning applies to the total fertility rates presented in section~\ref{calc_fert}.

\section{Statistical fluctuations}\label{stat_fluct}

Section~\ref{calc} dealt solely with the effects of artificially adding noise according to a given setup to the raw input counts.  A full analysis of the performance of such protection methods, however, should compare these effects to the \emph{intrinsic} statistical fluctuations that always affect these raw counts---and hence the resulting derived indicators---as well.  To obtain a model for these statistical fluctuations, it is a widely used practice in the literature to consider vital events as rare events during the reference period and thus as Poisson distributed; cf.\ e.g.~\citep{scott1981,peters2009}.  Then the Poisson parameter~$\lambda$, and hence the variance stemming from statistical fluctuation, of any given event count~$x_i$ can be estimated by the count itself: $\lambda=\text{Var}\simeq x_i$ thus giving
\begin{equation}\label{eq_delta_stat}
    \tilde{\Delta}_{i} = \sqrt{x_i}\,,
\end{equation}
where the tilde is to indicate that this uncertainty comes from intrinsic statistical variation (analog $\tilde{\delta}_i$ for relative uncertainties).  It can be verified with the data sets used in Section~\ref{calc} that this provides a robust estimation of the size of the intrinsic statistical variation.\footnote{
This can be done by calculating the standard deviation across the entire available time series for each NUTS~2 region, age and sex, and diving it by the estimator of Eq.~\eqref{eq_delta_stat}.  In the death counts, the median of this ratio over all available regions and ages is 1.2 (with lower and upper quartiles $1.0-1.7$) for males and 1.1 (lower/upper quartiles $0.9-1.6$) for females. In the live birth counts, the median is 1.8 (lower/upper quartiles $1.2-3.2$).  This shows that the variance estimator of Eq.~\eqref{eq_delta_stat} indeed provides a robust estimate of the magnitude of the statistical variation.  There is a systematic under-estimation of the actual variation over time (more so in the live births than in the deaths).  However, this is not surprising because the validation approach assumes constant event rates over the entire time series, which obviously underestimates the true variation.
}

Using Eq.~\eqref{eq_delta_stat}, all uncertainties calculated in Section~\ref{calc} can be modified or extended in a straightforward manner to reflect either statistical uncertainties or the combination of statistical and CKM noise uncertainties.  In all cases, the new expressions for combined uncertainties (indicated by a hat) result from replacing
\begin{equation}\label{eq_def_delta_hat}
    \Delta_{i}\rightarrow\hat{\Delta}_{i} = \sqrt{\Delta_i^2+\tilde{\Delta}_i^2} = \sqrt{\Delta^2+x_i}\,,
\end{equation}
For ease of reference,  we write out the modified Eqs.~\eqref{eq_delta_fx}, \eqref{eq_delta_f}, \eqref{eq_delta_Mx} and~\eqref{eq_delta_Ex_final}:
\begin{subequations}\label{eq_delta_hat}
\begin{align}
        \hat{\delta}_{f_x} &= \frac{1}{b_x}\sqrt{\Delta^2+b_x}\,; \label{eq_delta_fx_hat}\\
        \hat{\delta}_f &= \frac{1}{f} \sqrt{\sum_x \frac{\Delta^2+b_x}{w_x^2}}\,; \label{eq_delta_f_hat}\\
        \hat{\delta}_{M_x} &= \frac{1}{D_x}\sqrt{\Delta^2+D_x}\,; \label{eq_delta_Mx_hat}\\
        \hat{\delta}_{E_x} &= \frac{1}{E_x}\sqrt{\sum_{z=x}^{\overline{x}-1} \frac{\Delta^2+D_z}{B_z^2}\left(\frac{\ell_{z+1}}{\ell_x}\right)^2\left( \frac{1}{2}+E_{z+1} \right)^2 + (\Delta^2+D_{\overline{x}})\frac{B_{\overline{x}}^2}{D_{\overline{x}}^4} \left(\frac{\ell_{\overline{x}}}{\ell_x}\right)^2} \label{eq_delta_Ex_hat}\,.
\end{align}
\end{subequations}
In all cases, setting $\Delta=0$ produces a formula for the purely statistical uncertainty assuming the above-mentioned Poisson model.

\section{Combined analysis and NUTS~3 picture}\label{combined_analysis}

A full assessment of the effects of noisy SDC setups based on CKM or similar methods (the key property being a constant noise variance parameter) on the quality of demographic indicators must compare the total uncertainties ($\hat{\Delta}_i$, $\hat{\delta}_i$) to a scenario of only statistical uncertainties ($\tilde{\Delta}_i$, $\tilde{\delta}_i$), rather than comparing pure noise effects ($\Delta_i$, $\delta_i$) to a scenario of no uncertainties at all.  To this end, we make use of Eqs.~\eqref{eq_delta_fx_hat}--\eqref{eq_delta_Ex_hat} to analyse the relative admixture of CKM noise in the total uncertainty, calculated as $\hat{\delta}_i/\tilde{\delta}_i-1$ ($=\hat{\Delta}_i/\tilde{\Delta}_i-1$).  For the age-specific crude rates $f_x$ and $M_x$, Eqs.~\eqref{eq_delta_fx_hat} and~\eqref{eq_delta_Mx_hat}, there is a simple analytic expression as a function of the crude input event count that illustrates the limit in which the CKM admixture becomes negligible:
\begin{equation}\label{eq_tot_over_stat_limit}
    \frac{\hat{\delta_i}}{\tilde{\delta_i}} - 1 = \sqrt{1+\frac{V}{x_i}} - 1 = \frac{V}{2x_i} + \mathcal{O}\left( (V/x_i)^2 \right)\quad \text{for}\quad V\ll x_i\,.
\end{equation}
This makes explicit that relative CKM noise effects for these rates are small and thus negligible except where the CKM variance parameter is of the same order of magnitude as the crude event count, i.e.\ $x_i\sim\mathcal{O}(V)$.  However, since $V\sim\mathcal{O}(1)$ (e.g.\ $V\leq 5$) by design in all practical scenarios, this implies a regime where the statistical fluctuations are large, too, since $\tilde{\delta_i}=1/\sqrt{x_i}\sim\mathcal{O}(1/\sqrt{V})\sim\mathcal{O}(1)$.
\begin{figure}[t!]
\centering
\includegraphics[trim={0 40 0 30},clip,scale=0.335]{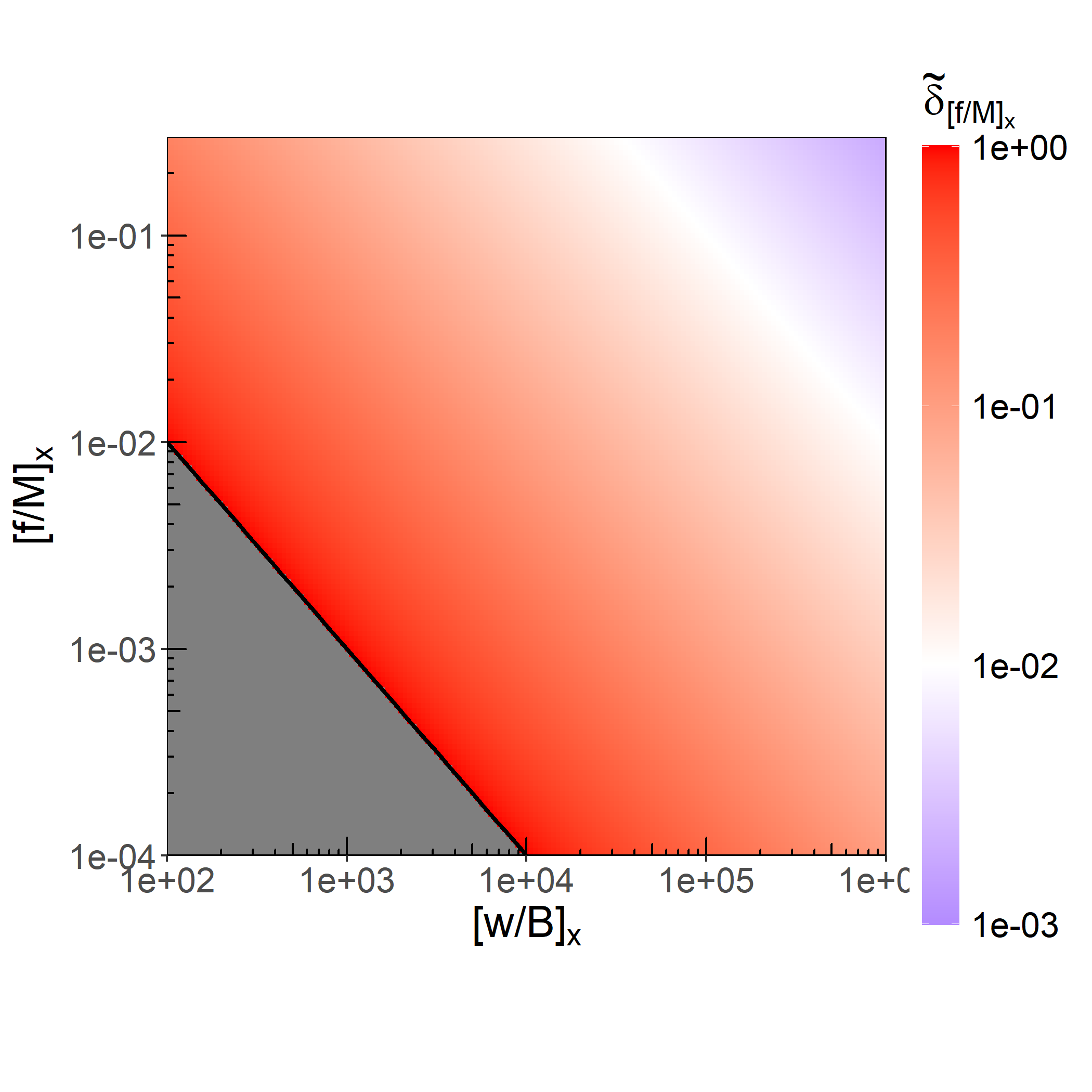}
\includegraphics[trim={0 40 0 30},clip,scale=0.335]{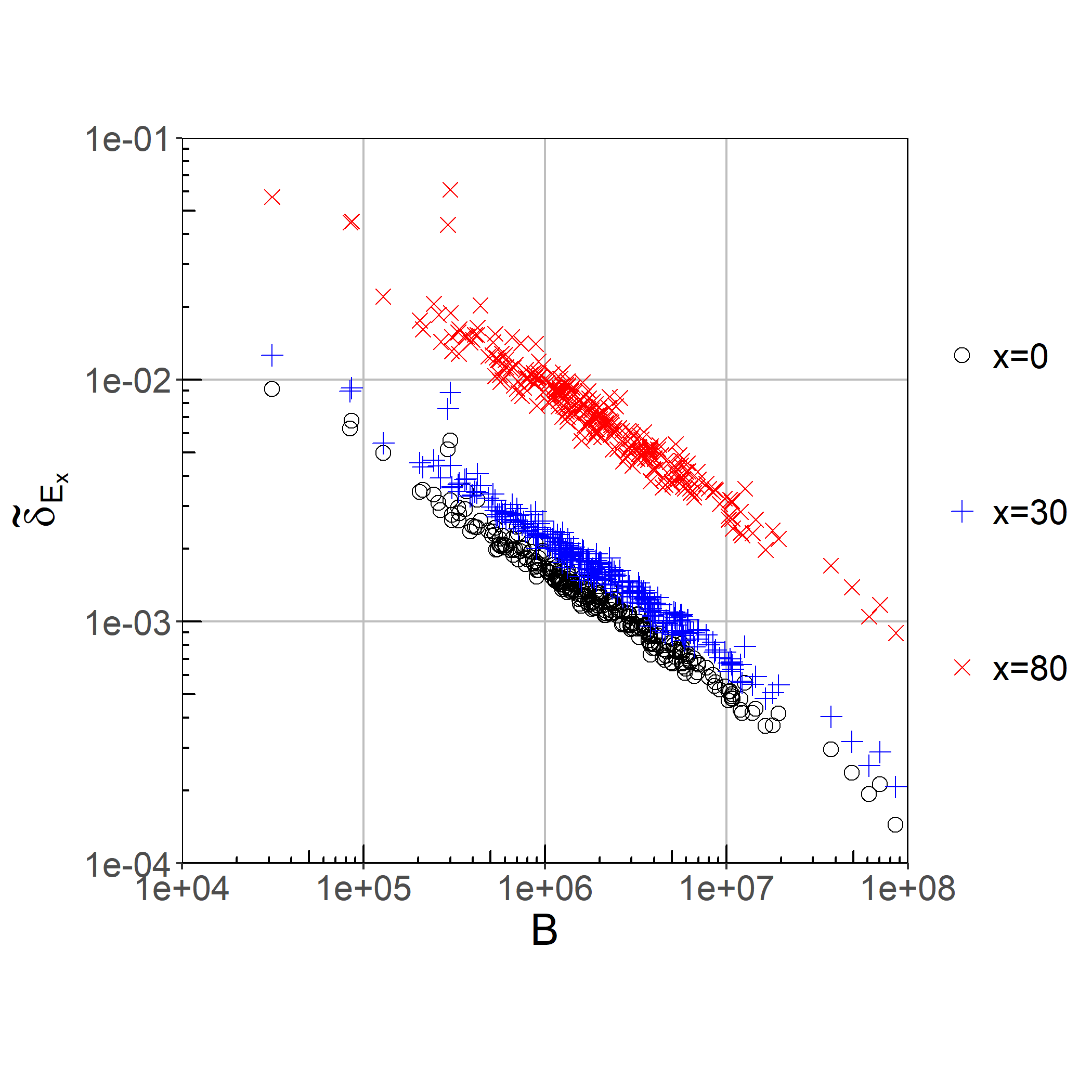}
\includegraphics[trim={0 40 0 30},clip,scale=0.335]{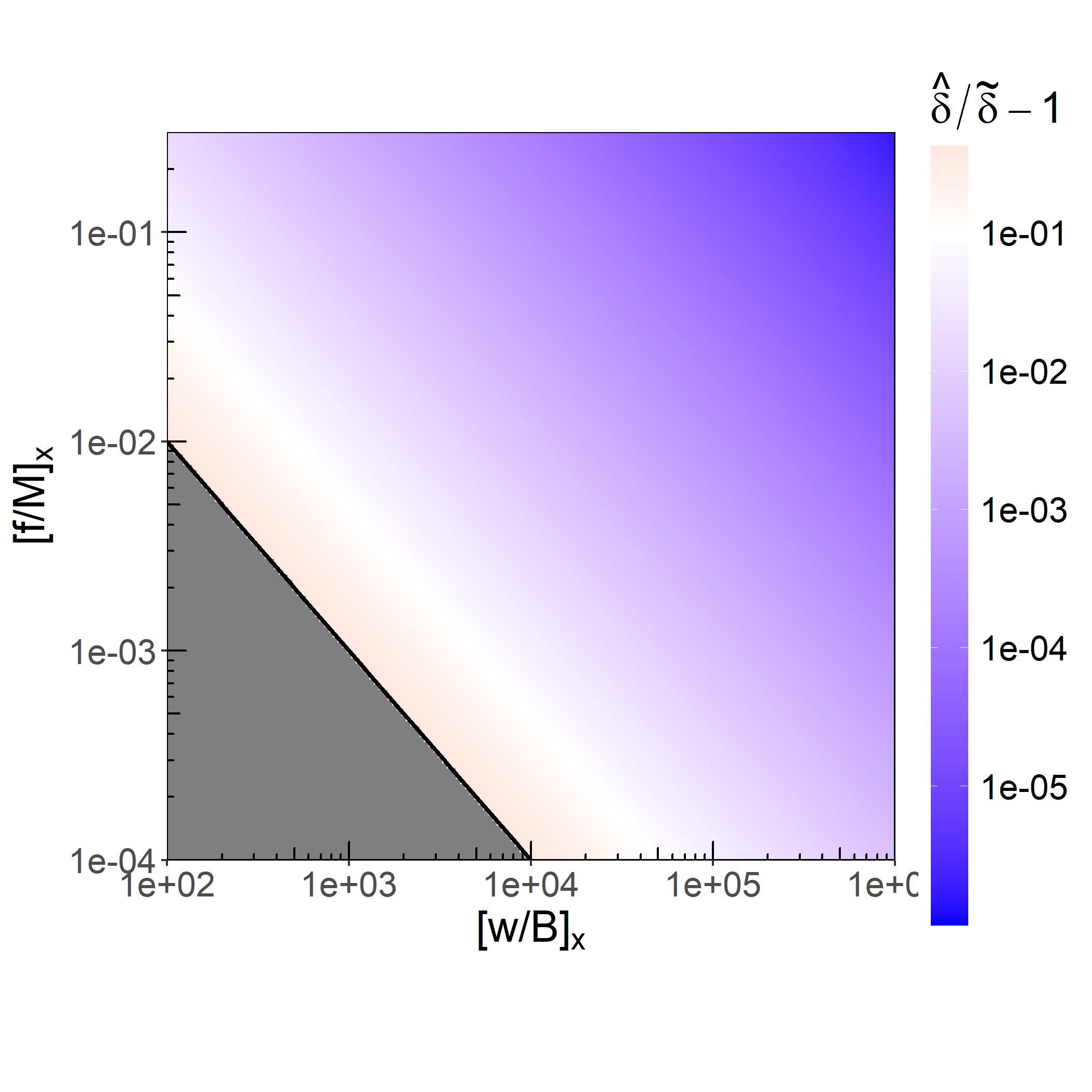}
\includegraphics[trim={0 40 0 30},clip,scale=0.335]{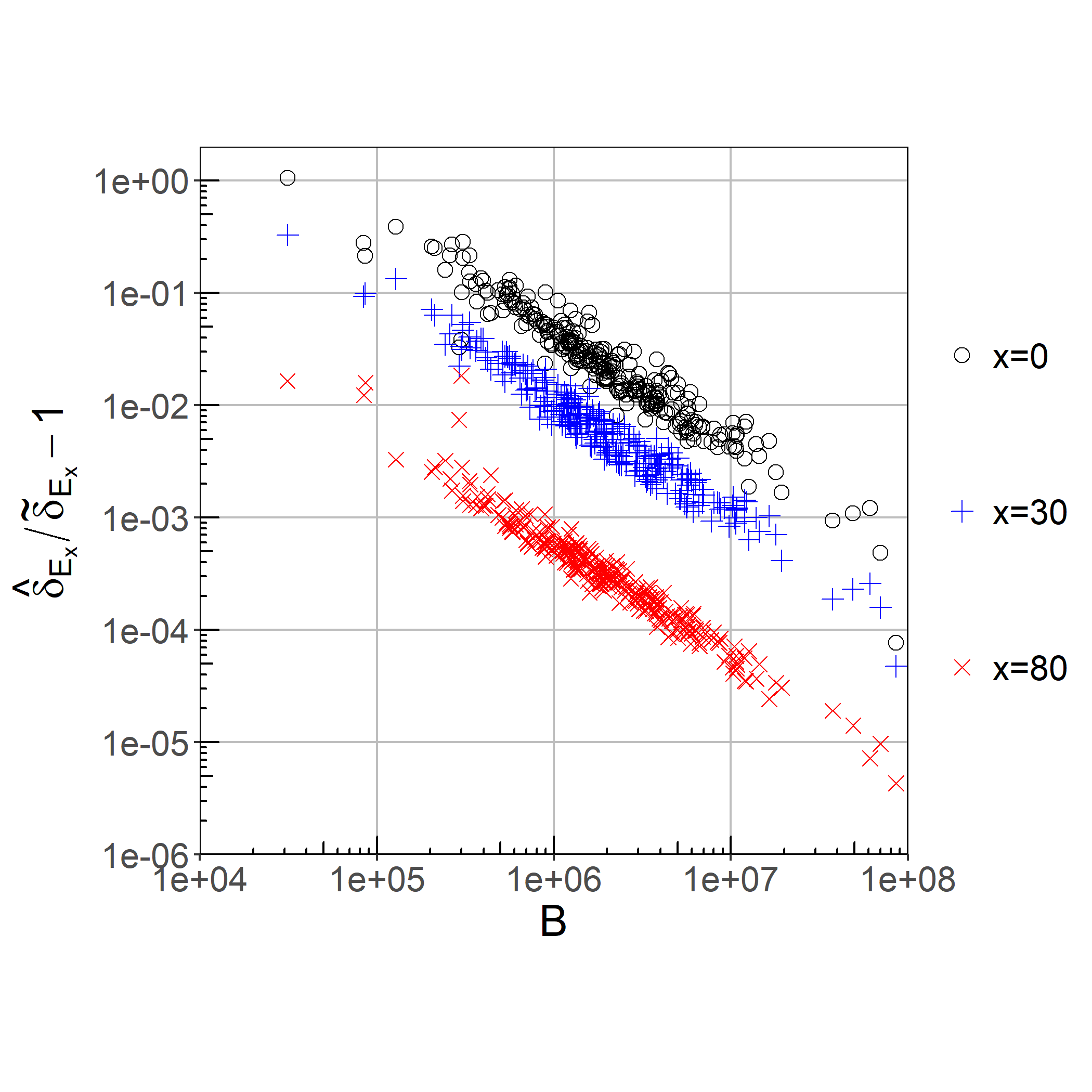}
\includegraphics[trim={0 40 0 55},clip,scale=0.335]{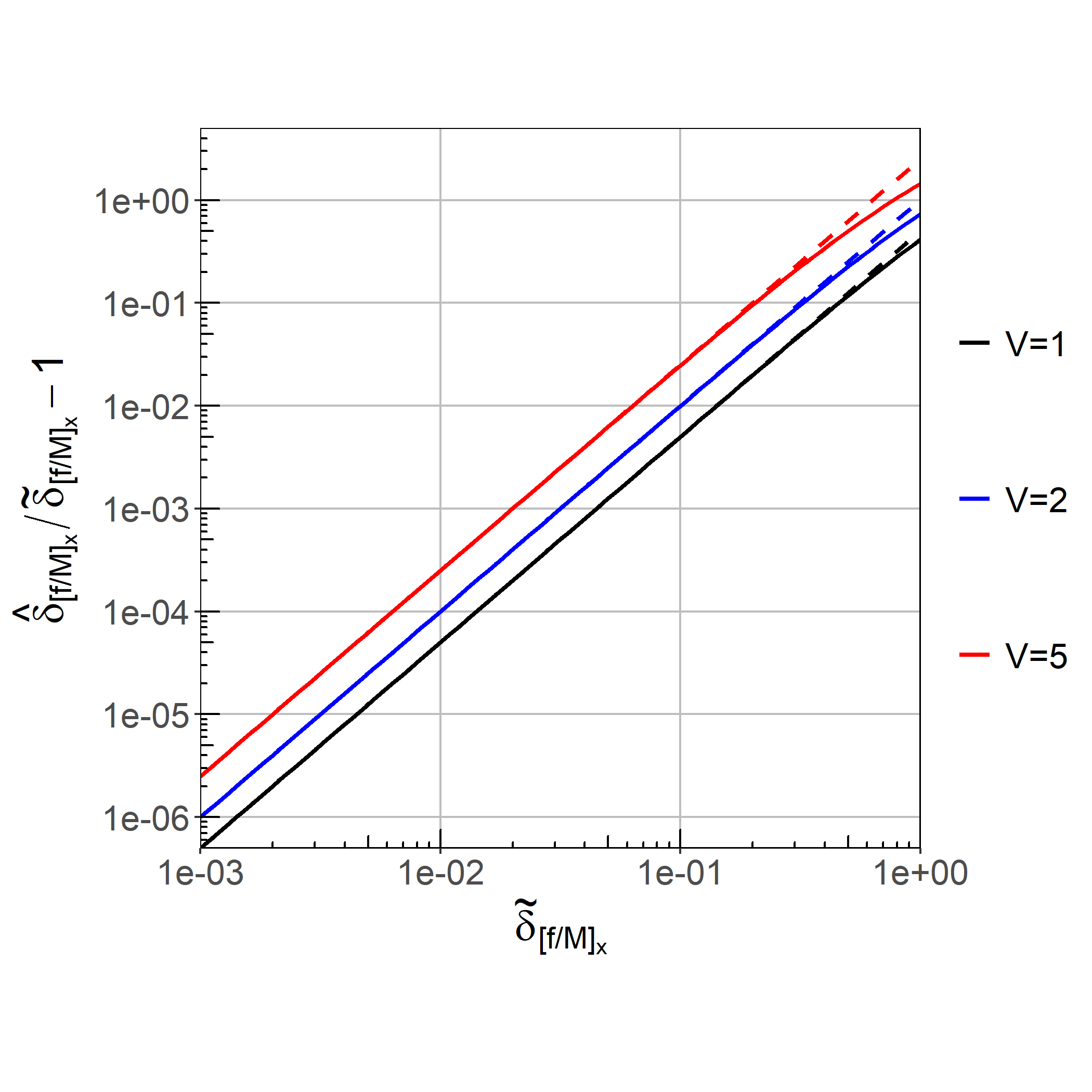}
\includegraphics[trim={0 40 0 55},clip,scale=0.335]{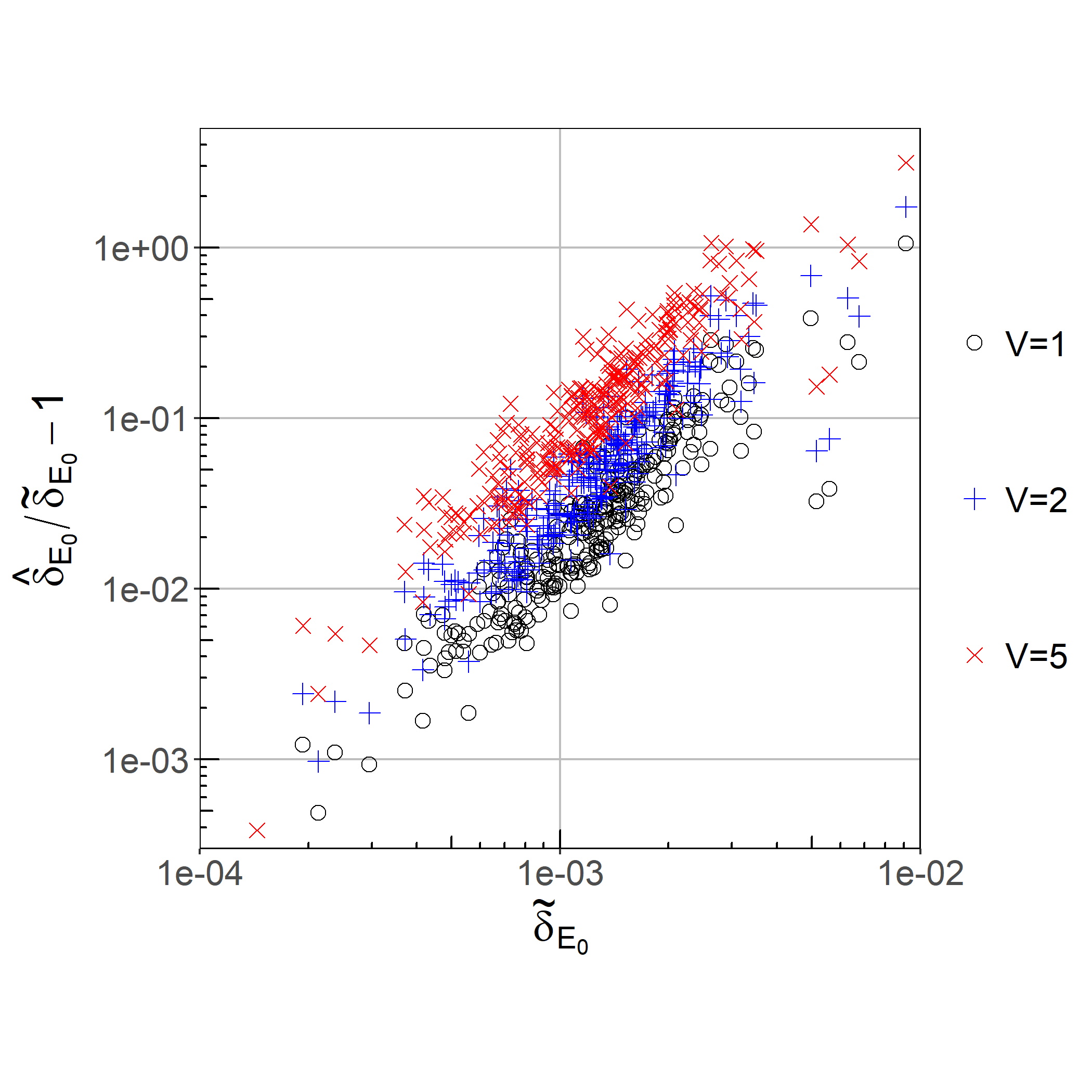}
\caption{Relative uncertainties from statistical fluctuations $\tilde{\delta}$ (top row) as well as relative admixture of CKM noise to the total uncertainty $\hat{\delta}/\tilde{\delta}-1$ against population size with $V=1$ (centre row) and against statistical uncertainty $\tilde{\delta}$ for various $V$ values (bottom row), as analytic plots for simple age-specific rates $f_x$ resp. $M_x$ (left column) and as scatter plots showing the 2023 NUTS~2 data for age-specific life expectancies (right column).  All results based on Eqs.~\eqref{eq_delta_fx_hat}--\eqref{eq_delta_Ex_hat} and~\eqref{eq_tot_over_stat_limit} (linear approximation dashed in the bottom--left plot).}
\label{fig_tot_over_stat}
\end{figure}

\begin{figure}[t!]
\centering
\includegraphics[scale=0.335]{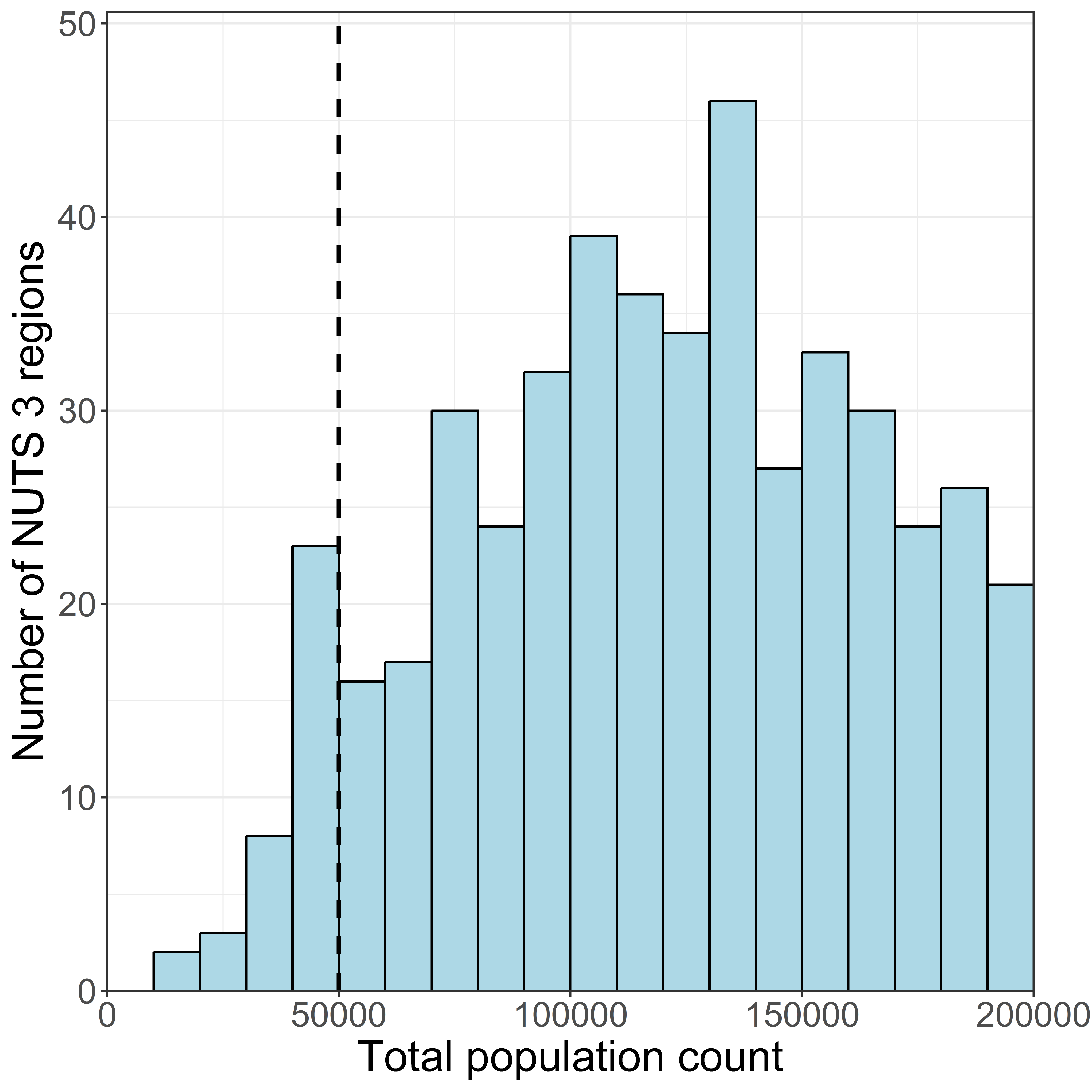}
\includegraphics[scale=0.335]{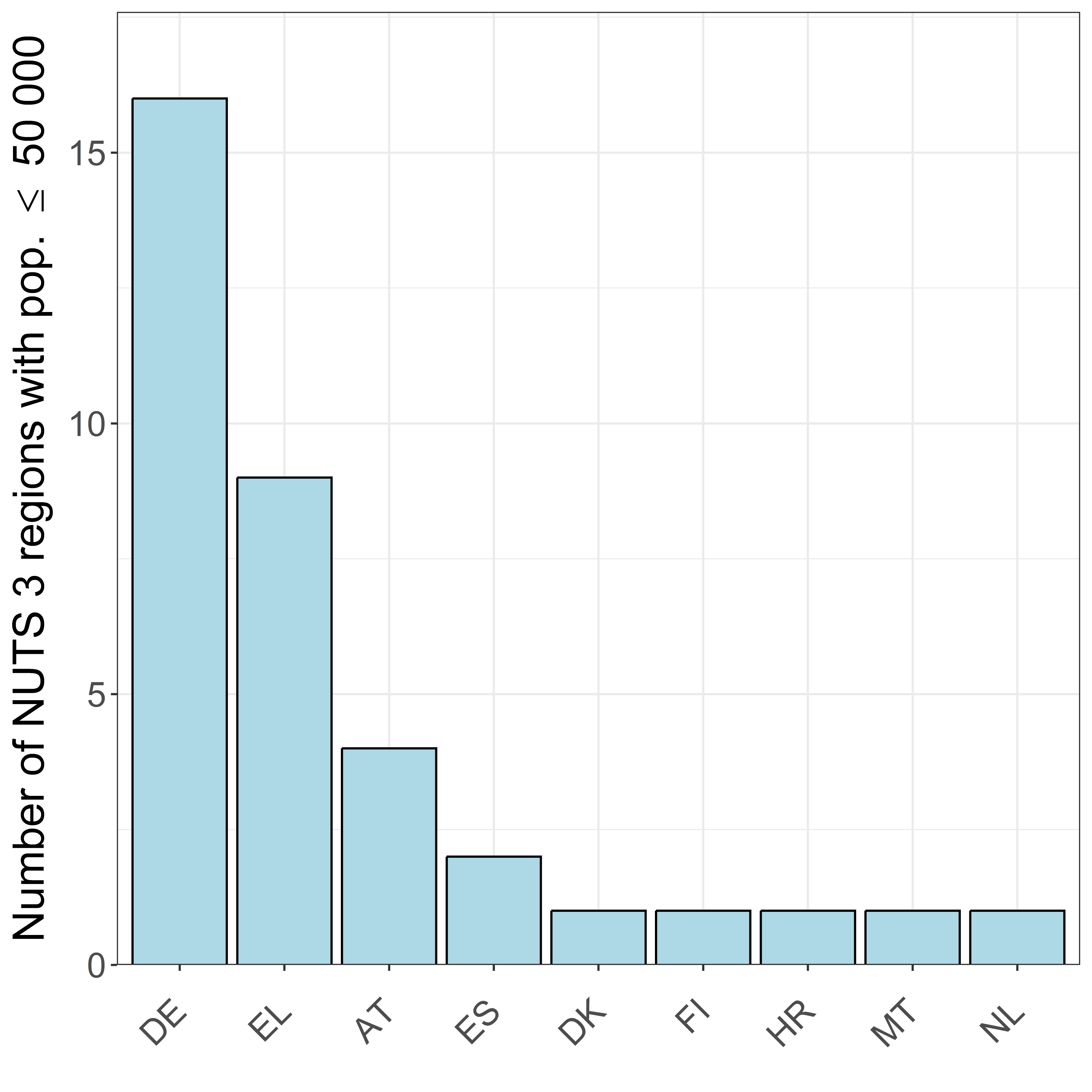}
\caption{Distribution of the total population of NUTS~3 regions on 1~Jan 2023 in the range up to $200\,000$ (left) and number of NUTS~3 regions with $\leq 50\,000$ people by EU country (right).}
\label{fig_histo_NUTS3}
\end{figure}

Fig.~\ref{fig_tot_over_stat} shows the comprehensive results for the age-specific fertility and mortality rates as well as for the life expectancies.  These results illustrate several key insights:
\begin{itemize}
    \item in general, statistical fluctuations~$\tilde{\delta}_i$ (assuming the Poisson model of Section~\ref{stat_fluct}) are an important source of uncertainty for all indicators considered here, with relative uncertainties $>1\,\%$ and up to $10\,\%$ for life expectancies in the higher age bands for populations $<100\,000$ people, and up to $100\,\%$ for age-specific event rates with underlying crude event counts of $\mathcal{O}(1)$;
    \item for the crude rates $f_x$ and $M_x$, the relative contribution of CKM noise to the total uncertainty~$\hat{\delta}_i$ remains negligible---i.e.\ $\hat{\delta}_i/\tilde{\delta}_i-1<10\,\%$---everywhere except in the extreme corners of the parameter space where the statistical fluctuations themselves become large, $\tilde{\delta}_i\sim\mathcal{O}(1)$, confirming the approximation of Eq.~\eqref{eq_tot_over_stat_limit};
    \item for the life expectancies~$E_x$, the relative contribution of CKM noise to the total uncertainty~$\hat{\delta}_i$ is generally larger than for the crude rates, with $\hat{\delta}_i/\tilde{\delta}_i-1\sim\mathcal{O}(100\,\%)$ for populations $<100\,000$ people and corresponding relative statistical fluctuations $\tilde{\delta}_i\sim\mathcal{O}(1\,\%)$;
    \item while increasing the CKM variance parameter within a realistic range $V\lesssim 5$ has a notable effect on the relative CKM noise contribution to the total uncertainty~$\hat{\delta}_i$, even the $V=5$ scenario does not generally lead to a situation where the CKM contribution becomes dominant for any indicator ($\hat{\delta}_i/\tilde{\delta}_i-1>300\,\%$ only for life expectancy at birth~$E_0$ of the smallest NUTS~2 region \r{A}land with a population of $\sim 30\,000$ people).
\end{itemize}
Finally, even though the data sets analysed here pertain to NUTS~2 level, these findings can be extrapolated to a NUTS~3 picture, due to the strong correlation of all uncertainties with population sizes that became explicit in the analysis above.  The population sizes of the NUTS~2 regions in the ESS are distributed mainly between $200\,000$ and $3\,000\,000$ people with a centre around $1\,000\,000$ people, whereas the NUTS~3 regions mainly lie between $10\,000$ and $500\,000$ people with a centre around $120\,000$ people; see Fig.~\ref{fig_histo_NUTS3}. There are no NUTS~3 regions with a population $<10\,000$ people and in fact only five regions $<30\,000$ people.
However, the regime of $30\,000$--$200\,000$ people is covered by four NUTS~2 regions, and thus included in principle in the analysis above.
%More specifically, all plots shown in Figs.~\ref{fig_fx_CKM}--\ref{fig_tot_over_stat} cover total populations $\gtrsim 10\,000$.
This implies that the findings would hold in principle for NUTS~3 level indicators, too, including life expectancies---even though the age-specific indicators and underlying crude event counts considered here are currently not published at NUTS~3 level.

%Due to the increased importance for any indicator release at NUTS~3 or other geographical levels with typically smaller population sizes (e.g.\ cities or functional urban areas),
It therefore merits to have a closer look at those few NUTS~2 regions with total population $<200\,000$ people, as these may be seen as archetypal for NUTS~3 level results, where the previous analysis suggests to focus on life expectancy at birth~$E_0$.  As detailed in Table~\ref{tab_LE_examples}, the  noise contribution to the total uncertainty becomes relevant in this regime ($\hat{\delta}/\tilde{\delta}-1\gtrsim 50\,\%$) even though statistical fluctuations remain very moderate ($\tilde{\delta}\lesssim 1\,\%$).  This means that in this regime, $E_0$ is a statistically very accurate indicator which however receives notable additional uncertainty from noise protection of the underlying raw counts.

At this point it must be stressed that some notable impact of noise protection on the resulting accuracy should be acceptable in principle.  After all, this merely exemplifies the fundamental trade-off between security and quality that is inherent to all disclosure control methods.  From this perspective, relative noise contributions up to $\hat{\delta}/\tilde{\delta}-1\sim\mathcal{O}(100\,\%)$ are not worrisome per se, especially if the noise protection allows in return to publish all raw counts and indicators without suppression. 

Nevertheless, Table~\ref{tab_LE_examples} shows the situation for a rather optimistic noise variance value $V=1$.  In practice, somewhat larger values in the range $V\lesssim 5$ should be more realistic.  However, setting $V=5$ brings relative noise impacts to $\hat{\delta}/\tilde{\delta}-1>100\,\%$ for all $E_0$ values in Table~\ref{tab_LE_examples}, and even to $400$--$700\,\%$ (depending on sex) for the smallest region \r{A}land (cf.\ lower-right plot in Fig.~\ref{fig_tot_over_stat}), so here the SDC noise arguably starts to dominate the total uncertainty.
Again, this should not be a reason to exclude noise protection but rather to apply particular care in analysing the effects in-depth and selecting an appropriate $V$~value as soon as population sizes $\leq 100\,000$ people come into focus.
\begin{table}[t!]
\centering
\resizebox{\textwidth}{!}{%
%\begin{tabular}{|l | c| c| c| c c| c c| c c| c|} % 11 columns
%    \hline
%    NUTS~2 region & Sex & $B$ & $E_0$ & $\Delta_{E_0}$ & $\delta_{E_0}$ & $\tilde{\Delta}_{E_0}$ & $\tilde{\delta}_{E_0}$ & $\hat{\Delta}_{E_0}$ & $\hat{\delta}_{E0}$ & $\hat{\delta}/\tilde{\delta}-1$ \\ %[0.5ex] 
%    \hline\hline
%    FI20 \r{A}land & T & $31\,225$ & 83.1 & $\pm 1.36$ & $1.64\,\%$ & $\pm 0.76$ & $0.91\,\%$ & $\pm 1.56$ & $1.88\,\%$ & $106\,\%$\\
%	   & M & $15\,412$ & 81.2 & $\pm 2.56$ & $3.15\,\%$ & $\pm 1.12$ & $1.38\,\%$ & $\pm 2.79$ & $3.43\,\%$ & $149\,\%$\\
%	   & F & $15\,813$ & 85.0 & $\pm 2.97$ & $3.49\,\%$ & $\pm 0.95$ & $1.12\,\%$ & $\pm 3.11$ & $3.66\,\%$ & $227\,\%$\\
%    \hline
%    ES63 Ceuta & T & $84\,349$ & 80.4 & $\pm 0.40$ & $0.50\,\%$ & $\pm 0.51$ & $0.63\,\%$ & $\pm 0.65$ & $0.80\,\%$ & $28\,\%$\\
%	   & M & $42\,316$ & 78.6 & $\pm 0.77$ & $0.97\,\%$ & $\pm 0.72$ & $0.92\,\%$ & $\pm 1.05$ & $1.34\,\%$ & $45\,\%$\\
%	   & F & $42\,033$ & 82.2 & $\pm 0.85$ & $1.03\,\%$ & $\pm 0.68$ & $0.83\,\%$ & $\pm 1.09$ & $1.32\,\%$ & $49\,\%$\\
%    \hline
%    ITC2 Aosta Valley & T & $128\,035$ & 82.4 & $\pm 0.39$ & $0.48\,\%$ & $\pm 0.41$ & $0.50\,\%$ & $\pm 0.57$ & $0.69\,\%$ & $39\,\%$\\
%	   & M & $61\,915$ & 80.7 & $\pm 0.74$ & $0.92\,\%$ & $\pm 0.54$ & $0.67\,\%$ & $\pm 0.92$ & $1.14\,\%$ & $70\,\%$\\
%	   & F & $66\,120$ & 84.1 & $\pm 0.83$ & $0.99\,\%$ & $\pm 0.63$ & $0.75\,\%$ & $\pm 1.05$ & $1.24\,\%$ & $66\,\%$\\ %[1ex] 
\begin{tabular}{|l | c| c| c| c c| c c| c c|} % 10 columns
    \hline
    & & & & & & \multicolumn{2}{|c|}{$V=1$} & \multicolumn{2}{|c|}{$V=5$}\\
    NUTS~2 region & Sex & $B$ & $E_0$ & $\tilde{\Delta}_{E_0}$ & $\tilde{\delta}_{E_0}$ & $\hat{\Delta}_{E_0}$ & $\hat{\delta}/\tilde{\delta}-1$ & $\hat{\Delta}_{E_0}$ & $\hat{\delta}/\tilde{\delta}-1$ \\ %[0.5ex] 
    \hline\hline
    FI20 \r{A}land & T & $31\,270$ & 85.1 & $\pm 0.70$ & $0.82\,\%$ & $\pm 1.59$ & $128\,\%$ & $\pm 3.27$ & $370\,\%$\\
	   & M & $15\,408$ & 82.0 & $\pm 1.09$ & $1.32\,\%$ & $\pm 2.86$ & $163\,\%$ & $\pm 6.01$ & $454\,\%$\\
	   & F & $15\,862$ & 88.5 & $\pm 0.82$ & $0.92\,\%$ & $\pm 3.25$ & $297\,\%$ & $\pm 7.09$ & $767\,\%$\\
    \hline
    ES63 Ceuta & T & $84\,310$ & 81.3 & $\pm 0.51$ & $0.62\,\%$ & $\pm 0.66$ & $30\,\%$ & $\pm 1.06$ & $110\,\%$\\
	   & M & $42\,294$ & 78.9 & $\pm 0.76$ & $0.96\,\%$ & $\pm 1.09$ & $44\,\%$ & $\pm 1.91$ & $154\,\%$\\
	   & F & $42\,016$ & 83.6 & $\pm 0.65$ & $0.77\,\%$ & $\pm 1.10$ & $70\,\%$ & $\pm 2.09$ & $224\,\%$\\
    \hline
    ITC2 Aosta Valley & T & $127\,856$ & 84.2 & $\pm 0.37$ & $0.44\,\%$ & $\pm 0.55$ & $49\,\%$ & $\pm 0.99$ & $166\,\%$\\
	   & M & $61\,898$ & 82.2 & $\pm 0.49$ & $0.59\,\%$ & $\pm 0.91$ & $87\,\%$ & $\pm 2.18$ & $266\,\%$\\
	   & F & $65\,958$ & 86.0 & $\pm 0.55$ & $0.64\,\%$ & $\pm 1.03$ & $87\,\%$ & $\pm 2.34$ & $268\,\%$\\ %[1ex] 
    \hline
\end{tabular}}
%\caption{Total population~$B$ and life expectancy at birth~$E_0$ by sex as well as absolute and relative uncertainty components, calculated with Eqs.~\eqref{eq_delta_Ex_final} resp.\ \eqref{eq_delta_Ex_hat} and $V=1$ for three selected small NUTS~2 regions.}
\caption{2023 Eurostat data on total population~$B$ and life expectancy at birth~$E_0$ by sex for three selected small NUTS~2 regions, shown together with uncertainties from statistical fluctuation alone ($\tilde{\Delta}_{E_0}$) as well as combined with CKM noise ($\hat{\Delta}_{E_0}$), calculated with Eqs.~\eqref{eq_delta_Ex_final} resp.\ \eqref{eq_delta_Ex_hat} for $V=1$ and $V=5$.}
\label{tab_LE_examples}
\end{table}

\section{Conclusions}

This paper computes and analyses the absolute and relative uncertainties on selected demographic indicators (age-specific fertility and mortality rates and life expectancies) when the raw input counts of births, deaths and population stocks are confidentiality protected by a generic noise method with fixed variance parameter~$V$.  We obtain analytical expressions by applying uncertainty propagation based on linear response, as well as numerical results using demographic data published by Eurostat.  To quantify the relative effect of noise protection, total uncertainties stemming from statistical fluctuations (according to a Poisson model) combined with additional noise uncertainties are compared to uncertainties only from statistical fluctuations.

The findings for combined relative uncertainties are that these become most pronounced in situations where the demographic indicator is dominated by very small raw input counts $<10$. This is the case for crude age-specific fertility rates, where relative uncertainties can thus increase up to $100\%$ for very small birth counts, and in particular for crude age-specific death rates, because small death counts are prevalent in many age bands.  Vice versa, the relative uncertainties remain moderate ($<10\,\%$) and are often even negligible ($<1\,\%$) for more aggregated indicators like total fertility rates and age-specific life expectancies.

Regarding the relative impact of noise uncertainty, the analysis has shown that this remains negligible---i.e.\ relative increase of total uncertainty compared to statistical fluctuation $<10\,\%$---for practically the entire relevant parameter space of all indicators analysed.  In general, noise effects become notable only in regimes where also the statistical fluctuations become large, even though the relative noise impact is bigger for life expectancies than for crude rates.
While most results are calculated with $V=1$, some additional variations with $V\leq 5$ indicate that these conclusions generally hold for the whole $V$~range that is relevant in practice.  However, for life expectancy at birth of small populations $\leq 100\,000$~people with still small statistical fluctuations $\sim 1\,\%$, the noise may start to dominate the combined uncertainty if rather large variances $V\sim 5$ are chosen.

Finally, even though the relative uncertainties on all indicators correlate inversely with the population size, the present NUTS~2 level analysis includes in principle ranges $\gtrsim 10\,000$ people so that the results are expected to also hold at NUTS~3 level or for other territorial typologies with a larger share of units with $10\,000$ to $200\,000$ people (e.g.\ cities or functional urban areas).  %Nevertheless,
%relatively small population sizes of $10\,000$--$100\,000$ people will be more frequent in such data, where 
It was shown that particular care needs to be taken here when fixing the noise parameter~$V$ appropriately to limit the impact on the total uncertainty of life expectancy at birth.

\bibliographystyle{unsrt}
\bibliography{references}

\pagebreak
\appendix

\section{Numerical validation of life expectancy uncertainty}\label{Delta_Ex_check}

\begin{figure}[t]
\centering
\includegraphics[trim={0 100 0 69},clip,scale=0.68]{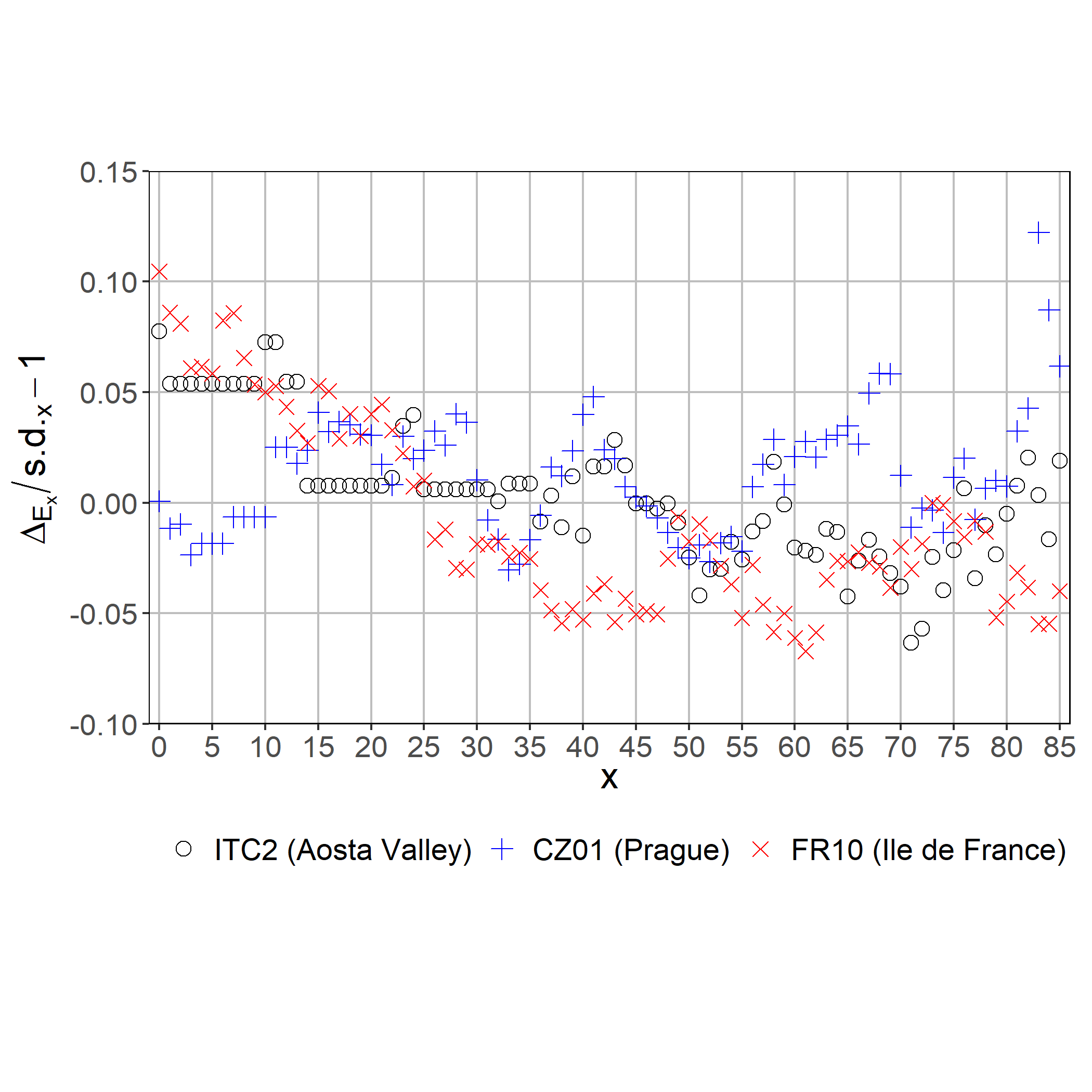}
\caption{Relative difference of statistical uncertainties from CKM noise as calculated analytically from Eq.~\eqref{eq_delta_Ex_final} ($\Delta_{E_x}$) and numerically from CKM noise sampling ($\text{s.d.}_x$), plotted over age~$x$ for the three selected NUTS~2 regions.}
\label{fig_Delta_Ex_check}
\end{figure}

Since the analytical derivation of Eq.~\eqref{eq_delta_Ex_final}, and thus of Eq.~\eqref{eq_delta_Ex_hat}, is not straightforward, we checked its correctness numerically as follows:
\begin{itemize}
    \item Three NUTS~2 regions were selected to represent typical population sizes: ITC2 (Aosta Valley) with $\sim 10^5$ inhabitants, CZ01 (Prague $\sim 10^6$) and FR10 (\^Ile de France $\sim 10^7$).
    \item For each region, 361 different versions of the crude age-specific death counts~$D_x$ (totals for reference year 2023 from {\it demo\_r\_magec}) have been sampled, each with a different simulated CKM noise protection added.\footnote{To draw the CKM noise, the R~package \hyperlink{https://cran.r-project.org/package=ptable}{{\it ptable}} was used with parameters $D=5$, $V=2$ and $j_s=2$.}
    \item For the original $D_x$~data and each CKM perturbed version, age-specific life expectancies~$E_x$ were calculated.
    \item From the set of 362~$E_x$ values for each $x\in \{0,\text{85+}\}$, the standard deviation ($\text{s.d.}_x$) originating from the CKM noise was estimated numerically from the statistical variation in the set, and then compared to the analytical result $\Delta_{E_x}$ as obtained from Eq.~\eqref{eq_delta_Ex_final}.
\end{itemize}
Fig.~\ref{fig_Delta_Ex_check} shows the relative differences $\Delta_{E_x}/\text{s.d.}_x-1$ against~$x$ for all three selected NUTS~2 regions.  It can be clearly seen that Eq.~\eqref{eq_delta_Ex_final} generally describes the uncertainty resulting from CKM noise added to the crude death counts very well, with deviations generally within $5\,\%$ and up to $10\,\%$ only for a small number of data points.  The correctness of the generalised uncertainty formula including also statistical fluctations, Eq.~\eqref{eq_delta_Ex_hat}, follows immediately, because the uncertainty generalisation step of Eq.~\eqref{eq_def_delta_hat} is analytically trivial.

One intricacy should be noted regarding the treatment of zeros in the $D_x$ data: in its standard setting, the CKM always leaves original zeros unperturbed.  This means that no variance is added from the zeros to the numerical $E_x$ sampling.  
However, Eq.~\eqref{eq_delta_Ex_final} does not generically differentiate between $D_x=0$ and $D_x>0$.  For the purpose of this validation, this can be easily lifted by requiring $D_z>0$ for the corresponding term to contribute to the sum in Eq.~\eqref{eq_delta_Ex}.  
If this fix is not applied, the analytical result tends to moderately overestimate the uncertainty, typically by less than a facor of~2 for low ages ($x<35$) in small populations $<200\,000$ people where $D_x=0$ data points are more abundant.  
Vice versa, the discrepancy becomes negligible for $x>35$ in general, and for any~$x$ in populations $\gtrsim10^6$ people.
However, for the main part of this paper, this CKM fix was {\it not} applied in general so as not to hard-code in the results a specificity of the CKM implementation chosen for numerical validation (namely to leave zeros unperturbed).  Moreover, the analytical uncertainties without the CKM fix always give the more conservative, i.e.\ larger, estimate.
Finally, the uncertainty from statistical fluctuations (Eq.~\eqref{eq_delta_Ex_hat} with $V=0$) is generally not affected, as $\Delta_{D_z}=\sqrt{D_z}=0$ automatically for $D_z=0$ in the Poisson model applied here.

\end{document}